\documentstyle[aps,psfig]{revtex}

\newcommand{\ket}[1]{|{#1}\rangle}
\newcommand{\bra}[1]{\langle{#1}|}

\newcommand{\slb}[2]{{#1}^{({#2})}}

\begin{document}
\twocolumn

\title{
A Cat-State Benchmark on a Seven Bit Quantum Computer
}

\author{
E. Knill$^1$(CIC-3), R. Laflamme$^1$(T-6) \\
R. Martinez$^1$ (CST-4), C.-H. Tseng$^2$
}
\address{ 
$^1$Los Alamos National Laboratory\\
$^2$Dept. of Nuclear Engineering, MIT, Cambridge, MA 02139
}
\maketitle


{\bf We propose and experimentally realize an algorithmic benchmark
that demonstrates coherent control with a sequence of quantum
operations that first generates and then decodes the cat state
$(\ket{000\ldots}+\ket{111\ldots})/\sqrt{2}$ to the standard initial
state $\ket{000\ldots}$.  This is the first high fidelity experimental
quantum algorithm on the currently largest physical quantum register,
which has seven quantum bits (qubits) provided by the nuclei
of crotonic acid.  The experiment has the additional benefit
of verifying a seven coherence in a generic system of coupled spins.
Our implementation combines numerous nuclear magnetic resonance (NMR)
techniques in one experiment and introduces practical methods for
translating quantum networks to control operations.  The experimental
procedure can be used as a reliable and efficient method for creating
a standard pseudo-pure state, the first step for implementing
traditional quantum algorithms in liquid state NMR. The benchmark and
the techniques can be adapted for use on other proposed quantum
devices. }

Quantum information processing (QIP) offers significant advantages
over classical information processing, both for efficient
algorithms~\cite{simon:qc1997a,shor:qc1995a} and for secure
communication~\cite{wiesner:qc1983a,bennett:qc1992a}. As a result it
is important to establish that sufficient and scalable control of a
large number of qubits can be achieved in practice.  There are a
rapidly growing number of proposed device
technologies~\cite{cirac:qc1995a,loss:qc1997a,bocko:qc1997a,shnirman:qc1997a,cory:qc1997a,chuang:qc1997a,privman:qc1997a,kane:qc1997a}
for QIP, and to compare them it is necessary to establish benchmark
experiments that are independent of the underlying physical system.  A
good benchmark for QIP should demonstrate the ability to reliably and
coherently control a reasonable number of qubits. This requires that
elementary operations can be implemented with small error regardless
of the state of the qubits, as sufficiently small error is one of the
most important prerequisites for robust
QIP~\cite{shor:qc1996a,aharonov:qc1996a,kitaev:qc1997a,knill:qc1998a}.
The cat-state benchmark proposed here is perhaps the simplest
demonstration of control which can be implemented for any number of
qubits and involves coherence in a non-trivial way.

To explain and realize the cat-state benchmark we use the example of
NMR based QIP.  At least two proposals
for quantum devices are based on using nuclear spins controlled by
radio frequency (RF) fields: The first involves the use of
molecules~\cite{cory:qc1997a,chuang:qc1997a} forming an ensemble of
quantum registers and the second uses nuclei embedded in a
semiconductor~\cite{privman:qc1997a,kane:qc1997a}. Of these proposals,
the first is presently accessible to experimental investigation by the
use of off-the-shelf equipment for liquid state NMR. By means of the
technique of preparing pseudo-pure states, it is possible to benchmark
quantum algorithms involving up to about ten qubits to determine how
well coherence is preserved and to measure how reliable the
available control methods are. There have been numerous experiments
implementing various quantum algorithms on up to five
qubits~\cite{jones:qc1998a,chuang:qc1998a,jones:qc1998b,chuang:qc1998c,chuang:qc1997b,laflamme:qc1997a,cory:qc1998a,nielsen:qc1998a,marx:qc1999a}
using NMR. The experiment reported here coherently implements a
quantum algorithm on seven qubits with a verifiable fidelity. It also
introduces a reliable method for preparing pseudo-pure states and for
verifying maximal coherences in generic spin systems.

\begin{figure}
\begin{center}
\mbox{\psfig{figure=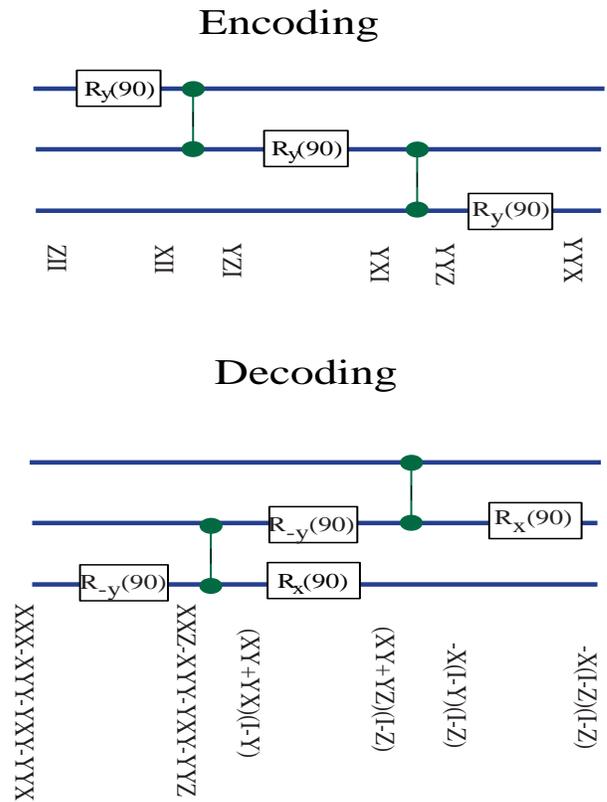,height=4.8in,width=3.2in}}
\end{center}
\caption{ Top: Encoding of the deviation matrix $\sigma_z II$ into
$\sigma_y\sigma_y\sigma_x$ by using a cascade of rotations and
J-couplings. A three coherence $\ket{000}\bra{111} +
\ket{111}\bra{000}$ is contained in the output which can be labeled
using a magnetic gradient or phase cycling.  Bottom: Decoding the
coherence to a pseudo-pure state is accomplished by a similar inverse
cascade.  The vertical text below the network denotes the state of the
three qubits at that point in the network with $X=\sigma_x,
Y=\sigma_y, Z=\sigma_z$.  Both networks generalize by extending the
cascade to more qubits.  }
\label{fig:3net}
\label{fig:3dnet}
\end{figure}

NMR QIP uses spin $1\over 2$ nuclei as qubits. Examples are protons and
carbon $13$ bound in a molecule.  QIP requires the ability to couple
different qubits. In molecules in a liquid at high magnetic field,
scalar couplings can be used for this purpose and controlled with
refocusing methods~\cite{cory:qc1997a,chuang:qc1997a}.  Thus, each
molecule can be considered as a quantum register consisting of (some
of) its spin $1\over 2$ nuclei. The initial state is prepared by
allowing enough time for thermal relaxation and readout is performed by an
ensemble measurement using standard NMR methods. We use deviation
density matrices~\cite{sorensen:qc1983a} for describing the state of
the nuclei.  To simplify the discussion, we use a three qubit example
of the cat-state benchmark. The thermal equilibrium state of a
molecule with one proton and two carbon $13$ nuclei at high field in a
liquid is given by $\slb{\sigma}{H}_z$ $+$ $.25\slb{\sigma}{C_1}_z$
$+$ $.25\slb{\sigma}{C_2}_z$ with high accuracy, up to an overall
scale factor and a multiple of the identity.  The standard Pauli
matrices are used as an operator basis, and superscripts on operators
refer to the particle the operator acts on. The cat-state benchmark
for this system begins by eliminating signal from the carbons to
obtain the initial state $\slb{\sigma}{H}_z$. Next a sequence of
quantum gates~\cite{barenco:qc1995a} is used to achieve the state
$\slb{\sigma}{H}_y\slb{\sigma}{C_1}_y\slb{\sigma}{C_2}_x$
(Fig.\ref{fig:3net}).  This state is a sum of several
coherences\cite{freeman:qc1998a}.  In particular, it contains the
three coherence $\ket{000}\bra{111}+\ket{111}\bra{000}$, which is the
deviation of the operator for the cat state
$(\ket{000}+\ket{111})/\sqrt{2}$. If each qubit is rotated by a phase
$\phi$ around the $z$-axis, the three coherence rotates by $3\phi$,
while all other components of this (or any other) state will rotate by
$0$, $\phi$ or $2\phi$. This feature can be used to label the three
coherence and eliminate all other components of the state, for example
by using a magnetic field gradient\cite{freeman:qc1998a}. An efficient
alternative using z-pulses or phase cycling is given below. The
three coherence can be decoded to the state
$\slb{\sigma}{H}_x\ket{00}\bra{00}$ (Fig.\ref{fig:3dnet}). This is
then observed after inverting the labeling gradient at three times the
original strength. In a fully resolved reference spectrum obtained by
applying a $90$deg rotation to the proton in the initial state, the proton
shows four peaks, one for each of the states $\ket{00}\bra{00}$,
$\ket{01}\bra{01}$, $\ket{10}\bra{10}$ and $\ket{11}\bra{11}$ of the
carbons. After decoding the cat state, only a single peak should be
left in the spectrum. The ratio $F$ of the intensity of this peak to
the intensity of the corresponding peak in the reference spectrum is
unity if everything works perfectly.  $F$ is reduced by errors in the
preparation and decoding steps.  Under the assumption that error in
the phase labeling method is negligible, it can be shown that $F$ is a
lower bound on the average of the fidelities~\cite{schumacher:qc1996a}
with which the decoding procedure maps the states
$\ket{000}\pm\bra{111}$ to the states $(\ket{0}\pm\ket{1})\ket{00}$.

\begin{figure}
\begin{center}
\mbox{\psfig{figure=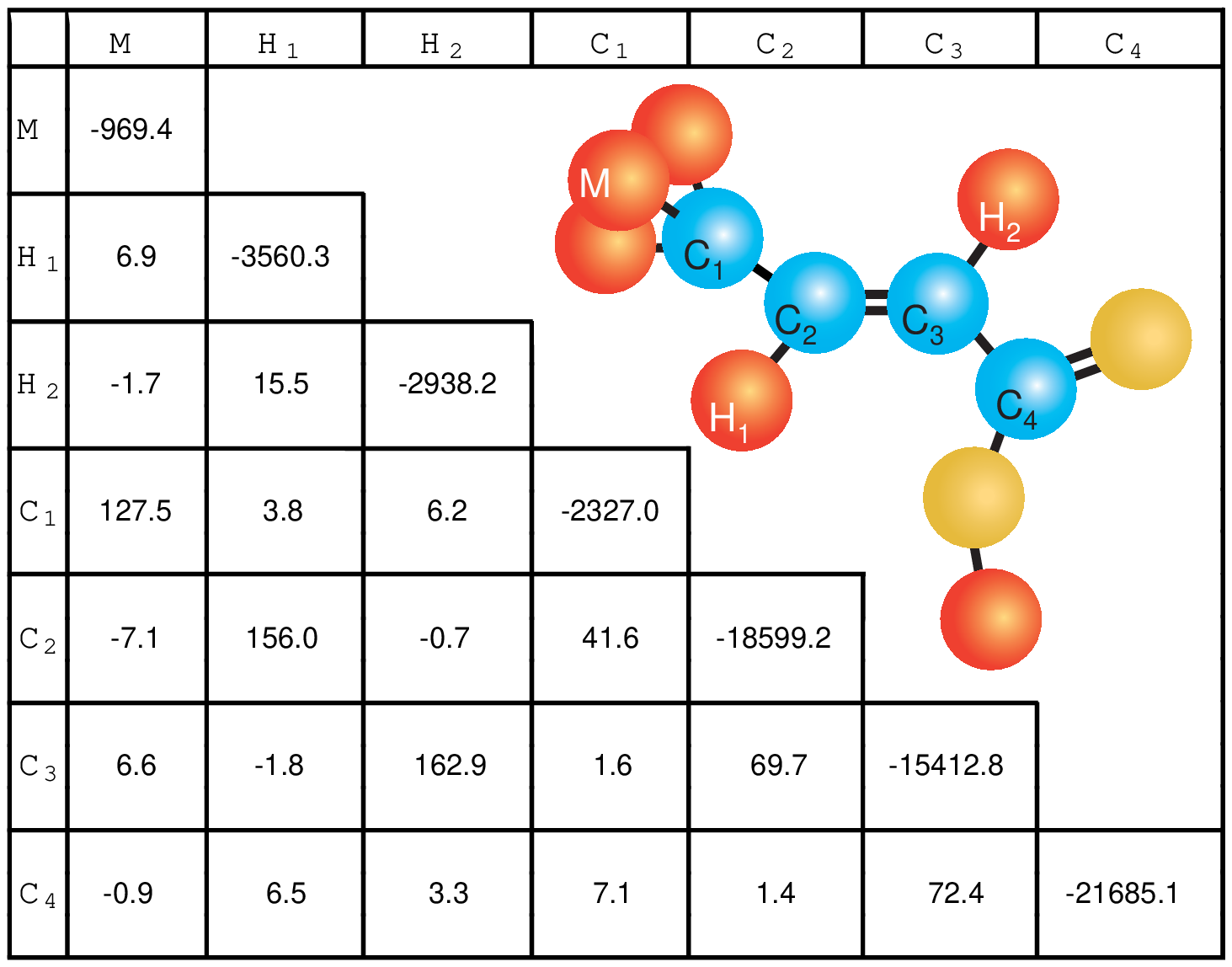,height=2.5in,width=3.2in}}
\end{center}
\caption{
Molecular structure of trans-crotonic acid together with a table of
the chemical shifts and $J$-coupling constants. The
chemical shifts are on the diagonal and are given with
respect to reference frequencies of $500.13$ MHz (protons)
and $125.76$ Mhz (carbons) on the $500$ Mhz spectrometer we used.
The T$^*_2$ were greater than $2$sec.
}
\label{fig:crot}
\label{tab:nucs}
\end{figure}

The three qubit cat-state benchmark can be generalized to any number
$n$ of qubits by repeating the steps of the cascade in the networks
shown in Fig.\ref{fig:3net}. We implemented the seven qubit version
using fully labeled trans-crotonic acid (Fig.\ref{fig:crot}). The
qubits are given by the spin $1\over 2$ component of the methyl group,
the two protons adjacent to the double bond and the four carbon $13$
nuclei.  A fidelity of $.73\pm.02$ was achieved. The loss of signal is
primarily due to spin relaxation, incomplete refocusing of couplings
and intrinsic defects in using selective pulses.  The success of the
experiment derives from the use of the following techniques: 1. An RF
imaging method to greatly reduce the effects of RF inhomogeneities.
2. A gradient based selection method for removing signal from the spin
$3\over 2$ component of the methyl group in an almost optimal way. 3.
The use of abstract reference frames for each nucleus to absorb
chemical shift and first order off-resonance effects in selective
pulses.  4. Precomputation of coupling effects during pulses.  5. A
pulse sequence compiler that optimizes delays between pulses for
achieving the desired amount of coupling evolution while minimizing
unwanted couplings.  All these techniques are scalable in
principle. Further details are in the methods section.

The cat-state benchmark has three applications that promise to make
it useful for NMR and other quantum technologies.  First, the
benchmark demonstrates the ability to reach the maximum
coherence with little loss of signal.  Previous experiments have generated
coherences by exploiting symmetry and effective Hamiltonian
methods. Very high order coherences can be observed in solid
state~\cite{pines:qc1994a}.  A maximal coherence of order seven was detected by
Weitekamp {\em et al.}~\cite{weitekamp:qc1982a} in benzene with one
carbon labeled by exploiting symmetry. The methods used in these cases
do not yield the amount of signal that can be achieved by using
methods based on quantum networks.

Second, the output of the benchmark can be used as a very reliable
pseudo-pure state for quantum algorithms.  We can write the maximum
coherence on $n$ qubits as a sum of two operators $\tilde X =
\ket{00\ldots}\bra{11\ldots}+\ket{11\ldots}\bra{00\ldots}$ and $\tilde
Y = i\ket{11\ldots}\bra{00\ldots}-i\ket{00\ldots}\bra{11\ldots}$.  The
decoding operator converts $\tilde X$ to
$\slb{\sigma}{1}_x\ket{0\ldots}\bra{0\ldots}$ and $\tilde Y$ to
$\slb{\sigma}{1}_y\ket{0\ldots}\bra{0\ldots}$.  These states can be
used as a pseudo-pure input to a quantum algorithm using one less
qubit, provided the following two problems are addressed: The first
problem is to ensure that the labeling method can be used together
with a subsequent algorithm.  The second problem is to eliminate
errors accumulated when decoding the $n$-coherence.

Clearly, the method used to label the $n$-coherence must be
reliable. The gradient based method is effective in conjunction with a
quantum algorithm, as long as the echo pulse is applied just before
the final observation.  Unfortunately, diffusion introduces loss of
signal at the gradient strengths required when used with long
algorithms.  Also, gradient methods do not easily generalize to other
proposed quantum devices--an important issue in benchmarking.  To
label the $n$-coherence one can instead perform $2n+1$ experiments,
where in the $k$'th experiment, the gradient is replaced by explicit
pulses that rotate each qubit by a phase $\phi_k=2\pi k/(2n+1)$.  If
$o_k$ is the expectation of the observable measured at the end of the
$k$'th experiment, then the value $o=\sum_k o_k e^{-i2\pi kn/(2n+1)}$
is non-zero only for signal originating at the $n$-coherence. This
technique can be applied in any system where it is possible to apply
$z$-rotations reliably. If the phase of applied pulses is highly
controllable (as is the case in systems controlled by RF or optical
fields), instead of applying explicit pulses to accomplish the
$z$-rotations, one can change the reference frame for each qubit, which
is equivalent to changing the phase of all subsequent pulses and the
observation reference phase by $-\phi_k$. This is essentially a phase
cycling method for selecting the $n$-coherence\cite{pines:qc1994a}.
We have used both the gradient based and this phase cycling method
with identical results in the crotonic acid system.

The problem of decoding error can in principle be solved by using the
maximum coherence directly as the input for a (modified) algorithm.
However it is not possible to obtain a reference signal for the
$n$-coherence without first mapping it to an accessible observable,
which can involve a loss of signal. Another problem is that it may be
inconvenient to use the $n$-coherence instead of the more familiar
standard pseudo-pure state. Our experiments show that we can decode the
$n$-coherence to the pseudo-pure state with no detectable error in the
observed spectrum. There can be error signal in unobserved operators
which one would like to eliminate from future observation.  This can be
done efficiently by performing multiple experiments, each with a
random phase of $0$deg or $180$deg applied to qubits $2,3\ldots$, a technique
which is a special case of the randomized methods of~\cite{knill:qc1997b}.
The number of experiments that need to be performed depends on
the desired level of suppression of possible error signals.
$N$ experiments result in suppression by a factor of $O(1/\sqrt{N})$.

The final application of the cat-state benchmark is as an experiment
to test the ability to coherently control a quantum system and
demonstrate a fully coherent implementation of a non-trivial quantum
algorithm.  This is a critical issue for scalable quantum information
processing, as scalable robustness requires that each operation has a
maximum error below some threshold (which may depend on the types of
errors)~\cite{aharonov:qc1996a,kitaev:qc1997a,knill:qc1998a}.  The
known thresholds seem to be dauntingly small. Nevertheless,
interesting small scale computations may be performable with much
higher error per gate. Thus the ability to implement the cat-state
benchmark with high fidelity is a good indication of what types of
tasks can be accomplished in the system at hand.  In addition, the
decoding algorithm of the cat-state benchmark is an instance of the
type of process required to perform fault-tolerant
error-correction~\cite{shor:qc1996a}, which is believed to be a
necessary subroutine in any large scale quantum computation. Our
experiment involved a total of twelve useful two-qubit operations, so
the fidelity of $.73$ suggests an error of about $.023=.27/12$ per
coupling gate. If this degree of control were available in the context
of quantum communication, it would be close to the known
thresholds~\cite{briegel:qc1998a}.

The realization of the cat-state benchmark given here is in an
ensemble setting. Most proposals for quantum devices involve individual
systems with pure initial states.  In these cases the benchmark can be
modified by replacing the ensemble measurements by repetition to infer
$o_k$ with sufficiently high signal to noise. The preparation step is
replaced by a network that directly maps the available initial state
to the cat state. Note that any evaluation of a quantum device
involves substantial repetition, essentially replacing the ensemble
measurement by an ensemble in time.

\noindent{\em Methods.} We used a Bruker DRX-500 NMR spectrometer with
a triple resonance probe for our experiments. (The triple resonance
probe is normally used for proton, carbon 13 and nitrogen 15; we used
only the first two.) All the equipment used was standard with no
specialized modifications.  The chemical structure of trans-crotonic
acid is given in Fig.\ref{fig:crot}. Deuterated chloroform
was used as the solvent. The chemical shifts (at 298K and
500Mhz) and coupling constants were experimentally determined to
within .1Hz by direct analysis of the proton and carbon spectra (see
Fig.\ref{tab:nucs}). This data was used to design selective pulse
shapes and times. Only $90$deg and $180$deg rotations were used in the
pulse sequence. The hard and selective pulses were analyzed by
simulation on single and pairs of nuclei and represented optimally as
a composition of phase shifts, $\sigma_z\sigma_z$ couplings and an
ideal $90$deg or $180$deg pulse. The simulation is efficiently scalable,
requiring $7(7+1)/2$ two qubit simulations for the seven qubit
register.  This permits elimination of most first order errors due to
off-resonance and coupling effects without using specialized shapes.
The computed phase shifts were absorbed into the rotating frame of
each nucleus, while the computed coupling effects contributed to the
coupling operations or were refocused.  To implement quantum
information processing tasks, we began with an ideal quantum network
expressed in terms of $90$deg rotations and $1/(2J)$ coupling
evolutions. Refocusing pulses are then inserted, and an optimizing
pulse sequence compiler is used to determine the best choice of delays
between pulses to achieve the desired evolution. The compiler
permitted us to automate many of the tasks of translating a quantum
network to a pulse sequence. Often smaller couplings cannot be
perfectly refocused without an excessive number of pulses.  Error due
to imperfect refocusing is explicitly estimated by the compiler. The
final pulse sequence used in our experiment required 48 pulses with an
estimated signal loss due to coupling errors of $.15$.

To have accurate pulses, we have reduced the effect of RF
inhomogeneities present in standard configurations by selecting signal
based on RF power.  The method first applies a $90$deg
excitation pulse to the methyl group followed by a sequence of pairs
of $180$deg rotations at phases of $\pm\phi_k$, where $\phi_k$ was
determined from one of the selective pulse shapes we used (modified by
an initial sequence to compensate for an off-resonance effect) and
designed to cause a $90$deg phase shift in the signal at an RF power of
about $\pm2\%$ of the ideal. At other powers, the effect is such that
a phase cycle involving a change of sign of the $90$deg phase shift
eliminates the signal.  A final pulse returns the selected signal
along the z-axis in preparation for the next step.  By calibrating the
power, we were able to retain $25\%$ of the signal compared to an
unselected spectrum. This sequence also has the property of selecting
signal from only the methyl group so that the initial state is
$\slb{\sigma}{M}_z$.

\begin{figure}
\begin{center}
\mbox{\psfig{figure=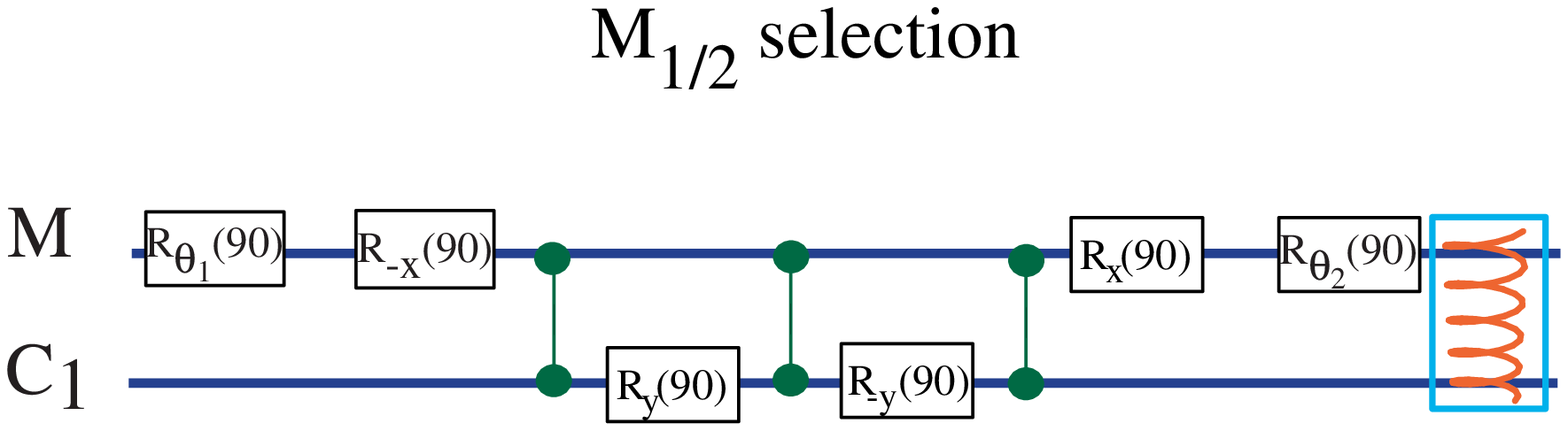,height=1.5in,width=3.2in}}
\end{center}
\caption{
Network to select the spin $1\over 2$ subspace of the methyl
group. Refocusings needed to decouple the other nuclei
are not shown. The spiral at the end is a gradient
``crusher'' to remove the transversal polarization.
}
\label{fig:msel}
\end{figure}

\begin{figure}
\begin{center}
\mbox{\psfig{figure=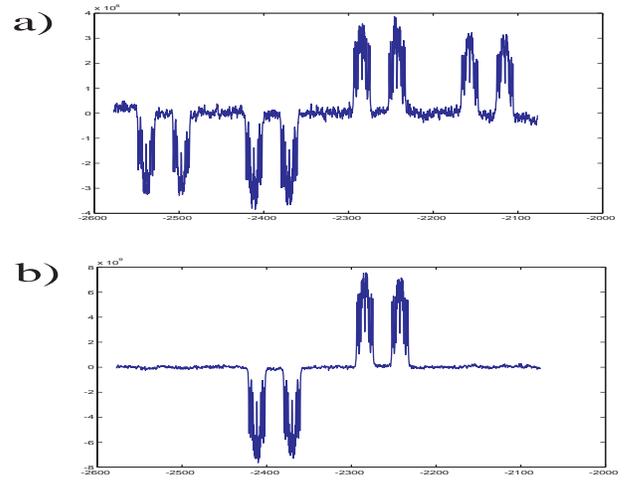,height=2.5in,width=3.2in}}
\end{center}
\caption{ Spectra of the methyl-carbon. Spectrum a) is obtained after
transfer of the methyl equilibrium polarization to the carbon.  Both
the spin $1\over 2$ and the spin $3\over 2$ components are
present. Only spin $3\over 2$ signal contributes to the extreme peak
groups. Spectrum b) is obtained after transfer of the spin $1\over 2$
selected polarization from the methyl group. There is no signal in the
extreme peak groups detectable above the noise. This together with two
other, similar spectra (not shown) obtained after different delays for
coupling demonstrates good selection of the spin $1\over 2$ component
of the methyl group. The scales for the two spectra are
different, the number of scans for a) and b) was 8 and 256
respectively. }
\label{fig:m1/2}
\end{figure}

The next step in the experiment required selecting the spin $1/2$
component of the methyl group's state space. This can be accomplished
by use of a three step sequence involving transfer of polarization
to the adjacent carbon and terminated by a gradient ``crusher''
(Fig.\ref{fig:msel}). The elimination of signal from the
spin $3\over 2$ states was verified by three experiments
involving observation of the signal on the adjacent carbon
after transfer of the methyl polarization with different delays
for coupling. One of the resulting spectra is shown in Fig.\ref{fig:m1/2}
with the standard reference spectrum. We were unable to detect
error signal above the noise.

\begin{figure}
\begin{center}
\mbox{\psfig{figure=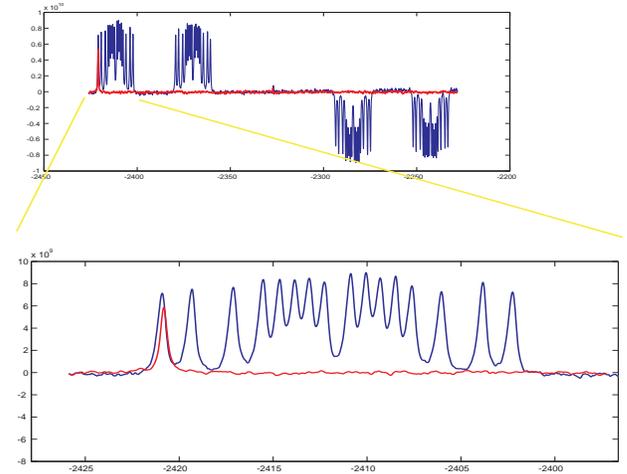,height=2.5in,width=3.2in}}
\end{center}
\caption{ Spectra of the pseudo-pure state (red) and the input state
after transfer of polarization to the methyl-carbon (blue).  Both
spectra were acquired with 256 scans and are shown at the same scale.
The signature of the pseudo-pure state is that only a single (the
leftmost) peak remains. No detectable signal remains in any other peak
position.  the intensity ratio for the left most peak in the
pseudo-pure state and the reference is $.73\pm .02$. .  }
\label{fig:pp}
\end{figure}

The remaining steps of the experiments consist of the generation of
the $n$-coherence, labeling the $n$-coherence, and the decoding
operations to obtain the standard pseudo-pure state, which was observed
on the methyl-carbon $C_1$. We chose $C_1$ for making observations
because all the couplings are adequately resolved there. The sequence
is as described earlier, with judiciously inserted refocusing pulses
and optimized delays. All the pulse phases were computed automatically
for the nuclei's individual reference frames. Knowledge of
the intended current state of a nucleus was exploited when that state
is $\ket{0}\bra{0}$ or $\ket{1}\bra{1}$ to absorb the effects of
couplings to that nucleus into the reference frame. The compiled pulse
sequence, pulse shapes and other required information needed for
running on a Bruker spectrometer is available from the authors.
Fig.\ref{fig:pp} shows the pseudo-pure state signal compared to a
reference spectrum obtained after selection of the spin $1\over 2$
selection sequence on the methyl group. Errors can show up as peaks in
positions different from the leftmost one.  We could not detect such
errors above the noise. The fidelity is given by the ratio of the
intensity of the left most peak in the final signal to the intensity
of a peak in the reference spectrum and was computed to be
$.73\pm.02$.

\noindent {\bf Acknowledgments.}  We thank C. Unkefer for help in
synthesizing labeled crotonic acid, David Cory and Tim Havel for
advice in using NMR spectrometers, S. Lacelle for suggesting the idea
of using crotonic acid, D. Lemaster and G. Fernandez for daily help at
the spectrometer and W.H. Zurek for encouraging us to exceed our
expectations.  This research was supported by the Department of Energy
under contract W-7405-ENG-36 and the National Security Agency.  We
thank the Newton Institute, where part of this work was completed.

\pagebreak

\newpage

\newpage

\end{document}